  \documentclass[traditabstract,referee]{aa} 
\usepackage{natbib,twoopt,amssymb,lscape,wasysym}
\usepackage[breaklinks=true]{hyperref} 
\bibpunct{(}{)}{;}{a}{}{,} 
\newcommandtwoopt{\citeads}[3][][]{\href{http://adsabs.harvard.edu/abs/#3}%
{\citealp[#1][#2]{#3}}} 
\newcommandtwoopt{\citepads}[3][][]{\href{http://adsabs.harvard.edu/abs/#3}%
{\citep[#1][#2]{#3}}} 
\newcommandtwoopt{\citetads}[3][][]{\href{http://adsabs.harvard.edu/abs/#3}%
{\citet[#1][#2]{#3}}} 
\newcommandtwoopt{\citeyearads}[3][][]%
{\href{http://adsabs.harvard.edu/abs/#3}{\citeyear[#1][#2]{#3}}}
\usepackage{graphicx}
\usepackage{longtable}
\usepackage{natbib}
\bibpunct{(}{),}{;}{a}{}{,} 
\usepackage{txfonts}
\newcommand{\chandra}{{\it Chandra}}

\newcommand{\suzaku}{{\it Suzaku}}

\newcommand{\swift}{{\it Swift}}
\newcommand{\integral}{\textit{INTEGRAL}}
\newcommand{\nustar}{\textit{NuSTAR}}

\newcommand{\ms}{$M_{\odot}$}

\newcommand{\fluxcgs}{ergs~s$^{-1}$~cm$^{-2}$}

\newcommand{\chisq}{$\chi^{2}_{\nu}$}

\begin{document}

   \title{X-ray, UV, and optical observations of the accretion disk and boundary layer in the symbiotic star RT~Crucis.}

   \author{G. J. M. Luna, 
          \inst{1,2,3,4}, 
          K. Mukai,
           \inst{5,6}
           J. L. Sokoloski,
           \inst{7}
           A. B. Lucy,
           \inst{7}
           G. Cusumano,
          \inst{8}
          A. Segreto,
          \inst{8}
          M. Jaque Arancibia,
          \inst{9,10}
          N. E. Nu\~nez,
           \inst{9}
          R. E. Puebla,
          \inst{11}
           T. Nelson, 
           \inst{12}    
           \and
           F. Walter,
           \inst{13}
           }

   \institute{CONICET-Universidad de Buenos Aires, Instituto de Astronom\'ia y F\'isica del Espacio, (IAFE), Av. Inte. G\"uiraldes 2620, C1428ZAA, Buenos Aires, Argentina\\
              \email{gjmluna@iafe.uba.ar}
                  \and
                  Universidad de Buenos Aires, Facultad de Ciencias Exactas y Naturales, Buenos Aires, Argentina
                  \and
                  Universidad Nacional Arturo Jauretche, Av. Calchaqu\'i 6200, F. Varela, Buenos Aires, Argentina
                  \and
                  INAF - Osservatorio di Padova, vicolo dell’ Osservatorio 5, I-35122 Padova, Italy
\and
CRESST and X-ray Astrophysics Laboratory, NASA Goddard Space Flight Center, Greenbelt, MD 20771, USA
\and
Department of Physics, University of Maryland, Baltimore County, 1000 Hilltop Circle, Baltimore, MD 21250, USA
\and
Columbia Astrophysics Lab 550 W120th St., 1027 Pupin Hall, MC 5247 Columbia University, New York, New York 10027, USA 
\and
INAF - Istituto di Astrofisica Spaziale e Fisica Cosmica, Via U. La Malfa 153, I-90146 Palermo, Italy
\and
Instituto de Ciencias Astron\'omicas, de la Tierra y del Espacio (ICATE-CONICET), Av. Espa\~na Sur 1512, J5402DSP, San Juan, Argentina
\and
Departamento de F\'isica y Astronom\'ia, Universidad de La Serena, Av. Cisternas 1200, La Serena, Chile.
\and
Facultad de Ingenier\'ia, Ciencias F\'isicas y Matem\'atica. Universidad Central del Ecuador. Alejandro Valdez y la Gasca / Ciudadela Universitaria. Quito-Ecuador.
\and
Department of Physics and Astronomy, University of Pittsburgh, Pittsburgh, PA 15260
\and
Department of Physics and Astronomy, Stony Brook University, Stony Brook, NY 11794, USA
 }
   \date{}

  \abstract
{Compared to mass transfer in cataclysmic variables, the nature of accretion in symbiotic binaries in which red giants transfer material to white dwarfs (WDs) has been difficult to uncover.  The accretion flows in a symbiotic binary are most clearly observable, however, when there is no quasi-steady shell burning on the WD to hide them.  RT~Cru is the prototype of such non-burning symbiotics, with its hard ($\delta$-type) X-ray emission providing a view of its innermost accretion structures.  In the past 20 years, RT Cru has experienced two similar optical brightening events, separated by $\sim$4000 days and with amplitudes of $\Delta$V$\sim$ 1.5 mag.  After \swift\ became operative, the Burst Alert Telescope (BAT) detector revealed a hard X-ray brightening event almost in coincidence with the second optical peak.

Spectral and timing analyses of multi-wavelength observations that we describe here, from \nustar, \suzaku, \swift/X-Ray Telescope(XRT) + BAT + UltraViolet Optical Telescope (UVOT) (photometry) and optical photometry and spectroscopy, indicate that accretion proceeds through a disk that reaches down to the white dwarf surface.  The scenario in which a massive, magnetic WD accretes from a magnetically truncated accretion disk is not supported.  For example, none of our data show the minute-time-scale periodic modulations (with tight upper limits from X-ray data) expected from a spinning, magnetic WD.  Moreover, the similarity of the UV and X-ray fluxes, as well as the approximate constancy of the hardness ratio within the BAT band, indicate that the boundary layer of the accretion disk remained optically thin to its own radiation throughout the brightening event, during which the rate of accretion onto the WD increased to 6.7$\times$10$^{-9}$ \ms\ yr$^{-1}$ (d/2 kpc)$^2$.  For the first time from a WD symbiotic, the \nustar\ spectrum showed a Compton reflection hump at E$>$ 10 keV, due to hard X-rays from the boundary layer reflecting off of the surface of the WD; the reflection amplitude was 0.77$\pm$0.21. The best fit spectral model, including reflection, gave a maximum post-shock temperature of $kT$=53$\pm$4 keV, which implies a WD mass of 1.25$\pm$0.02 \ms.

Although the long-term optical variability in RT~Cru is reminiscent of dwarf-novae-type outbursts, the hard X-ray behavior does not correspond to that observed in well-known dwarf nova.  An alternative explanation for the brightening events could be that they are due to an enhancement of the accretion rate as the WD travels through the red giant wind in a wide orbit, with a period of about $\sim$4000 days. In either case, the constancy of the hard X-ray spectrum while the accretion rate rose suggests that the accretion-rate threshold between a mostly optically thin and thick boundary layer, in this object, may be higher than previously thought.
}

\keywords{binaries: symbiotic -- accretion, accretion disks -- X-rays: binaries}
\authorrunning{G. J. M. Luna et al.}
\titlerunning{RT Crucis; X-ray, UV, and optical observations. }
\maketitle
%

\section{Introduction}

A binary system where a compact object accretes enough material from a red giant companion such that this interaction can be detected at some wavelength has been called a symbiotic system \citepads{2013A&A...559A...6L,2017arXiv170205898S}. If the compact object can be identified as a white dwarf (WD), then we refer to such systems as WD symbiotics. The orbital periods of symbiotics are of the order of a few hundred up to a few thousand days \citepads{2012BaltA..21....5M}. It is believed that mass transfer in most symbiotics proceeds via wind accretion; however, it might proceed via some form of wind Roche-lobe overflow. Podsiadlowski \& Mohamed (2007) proposed that in such an accretion scenario, even if the red giant does not fill its Roche lobe, its wind does. The wind is therefore focused toward the L1 point of the orbit, which increases both the likelihood of formation of an accretion disk around the white dwarf and the rate of mass flow into the white dwarf.  Other accretion modes were proposed to overcome the small efficiency of wind Roche-lobe overflow when the wind acceleration zone is not a significant fraction of the Roche lobe. \citetads{2015A&A...573A...8S} proposed that the wind from the giant in S-type symbiotics can be focused by its rotation, following the wind compression disk model from \citetads{1993ApJ...409..429B}. Unlike the smaller accretion disks in cataclysmic variable stars (CVs), when an accretion disk is formed in a symbiotic, its size should be about an AU or more \citepads{1986A&A...163...56D}. The symbiotic nebula, fed by the red giant' wind and ionized by the UV radiation field of the white dwarf (and possibly the accretion disk), can have densities on the order of a few 10$^{8-9}$ cm$^{-3}$. 

RT~Cru is the prototype of WD symbiotics with thermal, hard X-ray emission. Although a handful of systems are now known to have very hard X-ray emission similar to that of RT~Cru \citepads[see]{2009ApJ...701.1992K,2016MNRAS.461L...1M}, RT~Cru has unique characteristics. Usually, symbiotics get attention from observers when they experience some type of outburst, either "classical" or symbiotic novae-type \citepads[see][]{2016MNRAS.461.3599R} which are commonly detected first at optical wavelengths. On the other hand, RT~Cru got increased attention when it was first detected in the hard X-rays in 2003-2004 with \integral/IBIS \citepads{2005ATel..519....1C} at a $\sim$3 mCrab level. In 2012, INTEGRAL detected RT~Cru again, this time at a $\sim$13 mCrab level \citepads{2012ATel.3887....1S} and in 2015 at 6 mCrab level \citepads{2015ATel.8448....1S}. Since then, aside from multiple optical observations, RT~Cru has been observed with \chandra, \suzaku, \nustar\ and the \emph{Neil Gehrels Swift} Observatory \citepads{2004ApJ...611.1005G} XRT in X-rays and with \swift~ UVOT in UV. 

High temperature, hard-X-ray-emitting plasma is expected to be present if the white dwarf is accreting either via a disk or a magnetically channeled flow.  
The first X-ray observations of RT~Cru suggested that the high-energy emission originates in a highly absorbed multi-temperature optically thin thermal plasma perhaps from an accretion disk boundary layer instead of a magnetically channeled flow; a conclusion supported by the absence of periodic modulation in the X-ray emission. Subsequently, by equating the maximum temperature of the plasma with the shock speed, \citetads{2007ApJ...671..741L} were able to constrain the mass of the WD to be less than about 1.3 M$_{\odot}$.

If the hard X-rays are from a boundary layer, the inferred mass would put RT~Cru among those binaries hosting the most massive white dwarfs and become a firm candidate to be a supernova Type Ia progenitor. If the hard X-rays are from a boundary layer, we expect them to be reflected back into our line of sight by the WD surface and/or the accretion disk, producing a detectable Compton hump at energies of about 10 keV \citepads{1991MNRAS.249..352G}. If accretion is magnetic, then the X-ray emission from the accretion column would be reflected from the WD surface only, as the X-ray emitting column is far from the inner edge of the truncated accretion disk \citepads{2015ApJ...807L..30M}. Measurement of the Compton reflection -- as we report in this paper -- would constrain the geometry of the accretion flow and enable a precise determination of the mass of the WD.

In this paper we present the first reliable detection of the X-ray reflection component in a symbiotic star obtained with \nustar. This robust detection has allowed us to accurately determine the plasma shock temperature and thus constrain the WD mass to within 0.02 \ms. Additionally, we performed a comprehensive analysis of X-rays, UV, and optical observations of RT~Cru in order to shed light on the origin of the high-energy emission and the nature and timescale of the observed variability.  In Section \ref{sec:obs} we describe the observations and the reduction steps for every dataset analyzed. The results of the analysis of X-ray, UV, and optical data are described in Section \ref{sec:result}, while a discussion is presented in Section \ref{sec:disc} .In Section \ref{sec:concl}, we conclude that RT~Cru most likely accretes through a boundary layer rather than magnetic accretion columns, that the boundary layer is usually optically thin, and therefore that the boundary layer around a massive WD can remain optically thin up to surprisingly high accretion rates .

\section{Observations}
\label{sec:obs}

\begin{figure*}
\begin{center}
\includegraphics[scale=0.75]{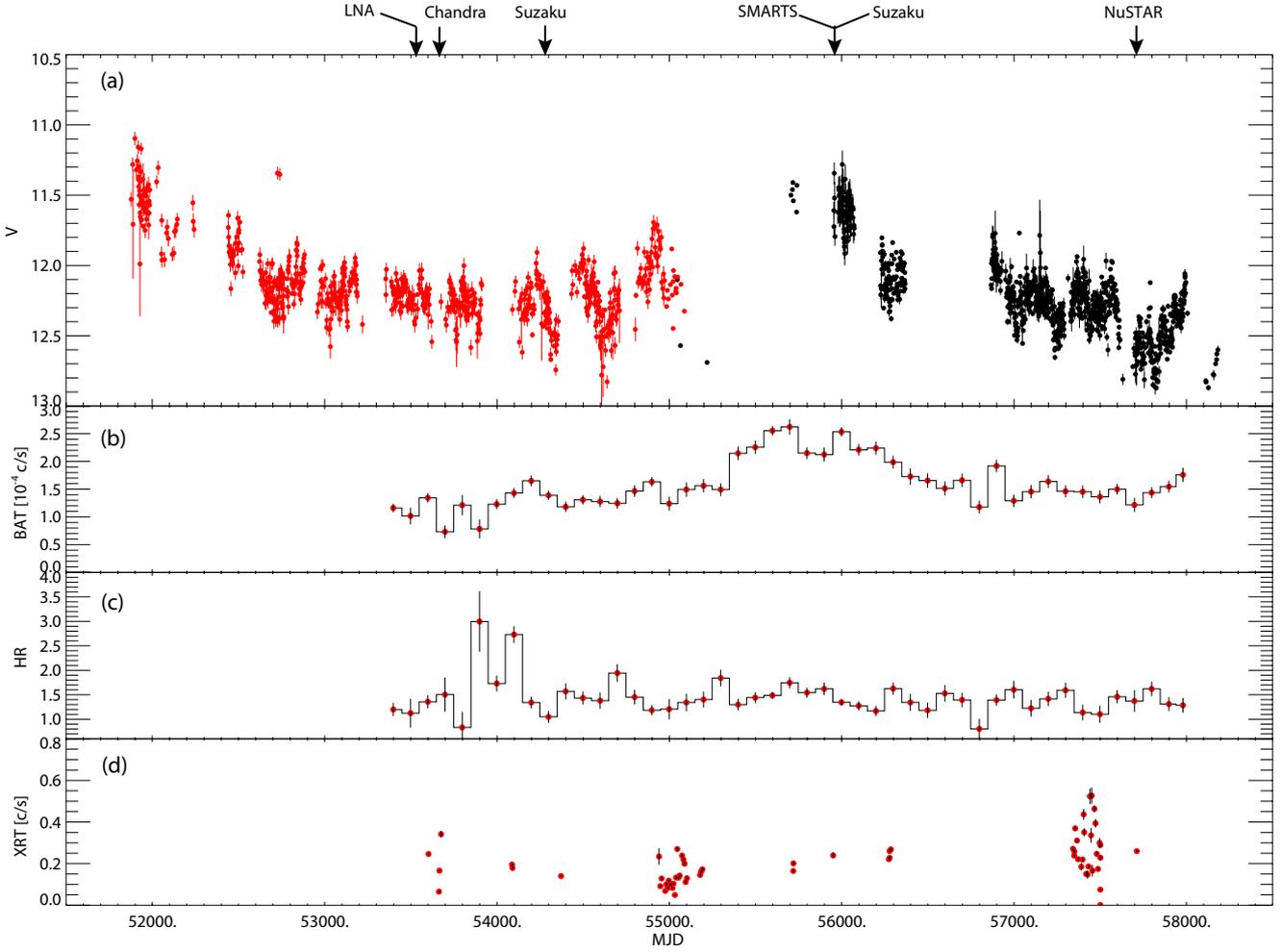}
\caption{{\em (a)} AAVSO (black dots) and ASAS (red dots) V-magnitude light curve. Arrows mark the dates when RT~Cru was observed with \chandra, \suzaku\, \nustar, LNA and SMARTS. {\em (b)} \swift~ BAT 153-months light curve with 100 days bin size in the energy range 15 to 85 keV; {\em (c)} \swift/BAT Hardness ratio (15-25/25-45 keV); {\em (d)}\swift/XRT light curve in the energy range 0.3-10 keV. }
\label{fig:aavso}
\end{center}
\end{figure*}

We observed RT~Cru with \nustar, \suzaku, \swift\ XRT+UVOT and in optical wavelengths with Laboratorio Nacional de Astrof\'isica (LNA) and Small \& Moderate Aperture Research Telescope System (SMARTS). We also collected multi-epoch photometry observations in the V-band from the American Association of Variable Star Observers (AAVSO) and the "All Sky Automated Survey" \citepads[ASAS]{2002AcA....52..397P} to study the X-ray data in the context of the optical state. The observations covered a period of 6\,019 days, since T$_{0}$= JD 2451870. For the ASAS data, we only considered those with quality flag "A" in the GRADE column and aperture to 30 arcsec. In the case of AAVSO, we ignored those data without measurement error bars. The upper panel in Figure \ref{fig:aavso} shows the resulting optical light curve.

\subsection{\nustar}
\label{sect:nustar}

\nustar~observed RT~Cru on 2016-11-22 for 58 ks. The data were reduced using the \nustar~ Data Analysis Software as part of HEASOFT 6.2 and filtered using standard filters given that the observation was not affected by abnormal solar activity. Using the tool \texttt{nuproducts,} we extracted source spectra and light curves from circular regions centered on the coordinates $\alpha$=12h 34m 53.74s, $\delta$=-64$^{\circ}$ 33$^{\prime}$ 56$^{\prime\prime}$ in the FPMB chip and $\alpha$=12h 34m 53.07s, $\delta$=-64$^{\circ}$ 33$^{\prime}$ 53.8$^{\prime\prime}$ in the FPMA chip \footnote{see \url{https://heasarc.gsfc.nasa.gov/docs/nustar/nustar\_faq.html\#AplusB}}, both with a 30$^{\prime\prime}$ radius. For the background, we chose an annular region with inner and outer radii of 110$^{\prime\prime}$ and 220$^{\prime\prime}$, respectively, and centered on the source coordinates. Light curves from the FPMA and FPMB cameras were summed and binned at 64 and 600 s for timing analysis.

\subsection{{\em Suzaku}}
\label{sect:suzaku}

We observed RT~Cru with \suzaku\ on 2007-07-02 (ObsID 402040010) and 2012-02-06 (ObsID 906007010). The data taken in 2007 (PI: J. Sokoloski) were already the subject of a publication by \citetads{2016A&A...592A..58D}. The second observation was triggered (through DDT time) in response to the announcement of a hard X-ray brightening detected with \integral~ \citepads{2012ATel.3887....1S}. The data obtained with \suzaku\ were reprocessed in order to apply the latest calibration available (v.2.2). We extracted the source spectra and light curves from the XIS detector by selecting photons from a circular region with a radius of 260 $^{\prime\prime}$ on each XIS chip. Four circular regions, each with 100$^{\prime\prime}$ radius, were used to extract the background (1: $\alpha$=12h 35m 29s, $\delta$=-64$^{\circ}$ 40$^{\prime}$ 31$^{\prime\prime}$; 2: $\alpha$=12h 34m 31.6s, $\delta$=-64$^{\circ}$ 41$^{\prime}$ 4$^{\prime\prime}$; 3: $\alpha$=12h 33m 49s, $\delta$=-64$^{\circ}$ 36$^{\prime}$ 37$^{\prime\prime}$; 4: $\alpha$=12h 35m 52s, $\delta$=-64$^{\circ}$ 35$^{\prime}$ 35$^{\prime\prime}$). Spectra from the XIS0 and XIS3 chips were combined using the \texttt{addascaspec} script. For the timing analysis, we combined the 3 XIS chips before extracting the light curves in order to increase the signal-to-noise ratio (S/N). Spectra and light curves from the HXD detector were extracted using the scripts \texttt{hxdpinxbpi} and \texttt{hxdpinxblc}, which take into account NXB and CXC backgrounds as well as dead-time corrections. Spectra were modeled using XSPEC while light curves, with a bin size of 64 and 600 s, were analyzed with custom codes.

\subsection{{\em Swift}}
\label{sect:swift}

\subsubsection{BAT survey detection}
\label{sect:bat}

To investigate the spectral behavior further, we retrieved the BAT survey data collected between 2005 January and 2017 August from the HEASARC public 
archive\footnote{http://heasarc.gsfc.nasa.gov/docs/archive.html} and processed them with the {\texttt{bat\_imager} code \citepads{2010A&A...510A..47S}, which performs screening, mosaicking, and source detection on data from coded mask instruments. For RT~Cru, we extracted light curves with a 100-days binning in three energy bands, 15--85, 15--25, and 25--45 keV,  also producing their hardness ratios (see Figure \ref{fig:aavso}.)

\subsubsection{Pointed XRT observations}
\label{sect:xrt}

As part of a monitoring campaign, during 2009, \swift\ observed RT~Cru for $\sim$3 ks every week for 6 months, for a total of 24 pointings and 72 ks. Other ToO observations during recent years have been requested in order to follow particular events such as \integral\ detections or quasi-simultaneous \chandra\ and \nustar\ observations. All the observations presented here used the photo counting (PC) mode of the XRT. We extracted source X-ray spectra and light curves from a circular region with a radius of 20 pixels ($\approx$47$^{\prime\prime}$) whose centroid we determined using the tool \texttt{xrtcentroid} and and found to be 0.3$^{\prime\prime}$ away from the SIMBAD coordinates. To correct for the presence of dead columns on the XRT CCDs during timing analysis of XRT data, we used the standard tool \texttt{xrtlccorr}. We extracted background events from an annular region with inner and outer radii of 25 and 40 pixels, respectively. We built the ancillary matrix (ARF) using the tool \texttt{xrtmkarf} and used the \texttt{swxpc0to12s6\_20130101v014.rmf} response matrix. 

\subsubsection{Pointed UVOT observations}
\label{sect:uvot}

During each visit with \swift\ we also obtained UVOT exposures in image and event mode. The UVOT observations in image-mode used either the V ($\lambda$5468 \AA, FWHM=769\AA), U ($\lambda$3465 \AA, FWHM=785 \AA), UVW1 ($\lambda$2600 \AA, FWHM=693 \AA), UVM2 ($\lambda$2246 \AA, FWHM=498 \AA), and/or UVW2 ($\lambda$1938 \AA, FWHM=657 \AA) filters while only the UVW2 filter was used with event-mode data.  Light curves from UVOT image-mode observations sample timescales of hours to days. In figure \ref{fig:rxte} we show the observation-averaged light curve from all the UVOT observations in the U, W1, M2 and W2 filters. Counts were converted to magnitudes and fluxes using the photometric zero points listed in \url{https://swift.gsfc.nasa.gov/analysis/uvot_digest/zeropts.html}. We used the E(B-V)=0.374 derived in Sect. \ref{sect:opt_spec} to correct for reddening. 

\begin{figure}
\includegraphics[scale=0.45]{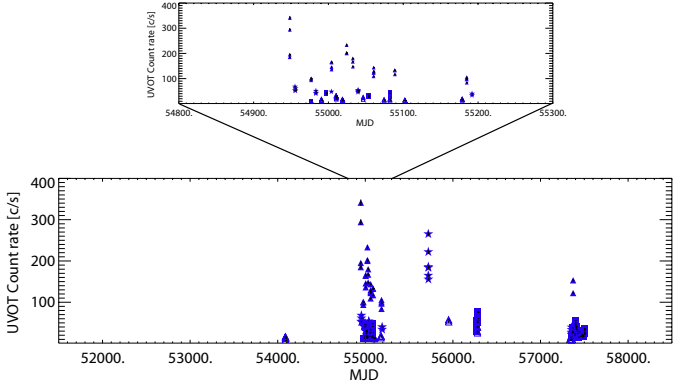}
\caption{Observation-averaged \swift\ UVOT light curve (U:filled $\triangle$; W1:$\star$; M2: open $\triangle$; W2: $\Square$). The upper panel shows a zoom from MJD 54800 to MJD 55300, when the majority of \swift\ observations were taken and strong variability was detected.}
\label{fig:rxte}
\end{figure}

Some intervals during pointed observations were affected by unstable spacecraft attitude. In cases where only image-mode was taken, we excluded those snapshots from our light curve analysis and extracted the source count rate (using the \texttt{uvotsource} script) within each snapshot from a circular region of 5$^{\prime\prime}$ radius and background from an annular region of 10$^{\prime\prime}$ and 20$^{\prime\prime}$ inner and outer radii, respectively. 

Unfiltered event files are distributed when event mode is requested. We filtered the event files using the \texttt{uvotscreen} tool\footnote{Information about satellite attitude was extracted from the corresponding \texttt{.mkf} file, while the filtering expression used was: \texttt{aoexpr=(SAC\_MODESTAT / 32) \% 2 == 1 $\&\&$ SETTLED==1 $\&\&$ ANG\_DIST $<$ 100. $\&\&$ ELV $>$ 10. $\&\&$ SAA == 0 $\&\&$ SAFEHOLD==0 $\&\&$ SAC\_ADERR$<$0.2 $\&\&$ STLOCKFL==1 $\&\&$ STAST\_LOSSFCN$<$1.0e-9 $\&\&$ TIME\{1\} - TIME $<$ 1.1 $\&\&$ TIME - TIME\{-1\} $<$ 1.1}} excluding those events taken during unstable spacecraft attitude (\texttt{SETTLED=1}). With the cleaned event file ready, which most of the time consists of a few snapshots, we first extracted one event file per snapshot and created snapshot images on which we used the 
\texttt{uvotevtlc} script to extract light curves with a bin size of 10 s, corrected by coincidence lost, large-scale sensitivity map and detector sensitivity. Figure \ref{lc:uvot-event} shows an example of such light curves extracted from one snapshot during ObsID \#00030840019. In order to extract information about the source variability in short time scales, we used custom codes to derive statistical quantities on each light curve that lasted more than 100 s.

\begin{figure}
\includegraphics[scale=0.5]{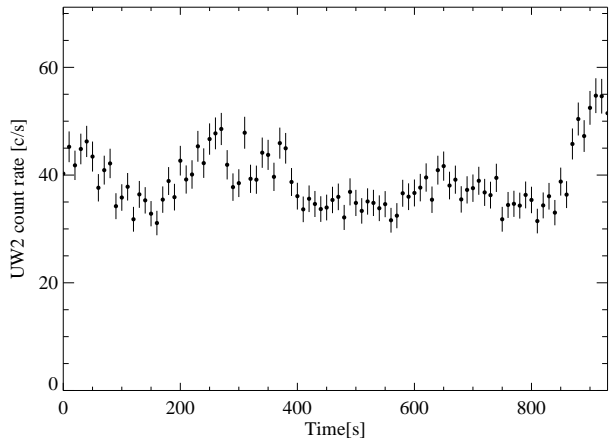}
\caption{{\em Swift}/UVOT event-mode light curve derived from the longest snapshot during ObsID \#00030840019 with a bin size of  10 s.}
\label{lc:uvot-event}
\end{figure}

\subsection{Optical spectroscopy}
\label{sect:opt_spec}

Low-resolution optical spectra were obtained using the 1.6m telescope in the Laboratorio Nacional de Astrof\'isica (Brazil) with the Cassegrain (with a dispersion of 4.4\AA/pixel) on 2005-06-11. The database\footnote{\url{http://www.lna.br/~databank/databank.html}} from this observatory also provided us with a few historical spectra from 2001-03-23, 2007-07-23 and 2018-02-09. We reduced the data using the IRAF\footnote{IRAF is distributed by the National Optical Astronomy Observatory, which is operated by the Association of Universities for Research in Astronomy, Inc. under cooperative agreement with the National Science Foundation.} package and followed the standard procedures of bias image subtraction, flat-fielding, and wavelength calibration. Flux calibration of the low-resolution spectra was secured through observations of standard stars on each night. Emission line fluxes were measured by fitting Gaussian profiles and a Gaussian deblending routine was also used when necessary. Reddening was derived using the Balmer emission lines fluxes and the method outlined in \citetads{2005A&A...435.1087L}, which was based on \citetads{1996PASP..108..972G}. The method derives the reddening by using the Balmer decrement values determined by \citetads{1975MNRAS.171..395N} under self-absorption conditions, appropriate for symbiotics. Given that symbiotics, in general, exhibit variability in their emission lines, our derivation of reddening should be taken as an order of magnitude estimation.

Low-dispersion spectra were also obtained with the SMARTS 1.5m/RC spectrograph in early 2012. We obtained blue spectra 3650-5415\AA\ with 4.1 \AA/pixel resolution on the night of 2012-02-05 (UT) and red spectra (3.1\AA/pixel resolution; 5630-6940\AA) on the night of 2012-02-08. All spectra are the sum of three spectra taken to filter cosmic rays. Total integration times were 1200 s on 5 February and 360 s on 8 February. Data reductions followed the methods described by \citetads{2012PASP..124.1057W}. 

\section{Results}
\label{sec:result}

\subsection{Long-term variability}

The historical optical light curve shows two similar brightening events separated by $\sim$4000 days. Almost in coincidence with the second event, with \swift\ being operative, there was an increase in the hard X-ray flux, as measured with BAT, of at least a factor of two (see Figure \ref{fig:aavso}).  Interestingly, as the hard X-ray flux increased throughout the major optical brightening, the hardness ratio within the BAT band remained fairly constant (see Figure \ref{fig:aavso}).  Moreover, \suzaku\ observations show that at energies below $\sim$ 4 keV --- where absorption has the strongest effect on X-rays, the X-ray flux decreased as the optical emission rose
(Figure \ref{fig:aavso}). The ratio of UV-to-X-ray flux grew during several optical bright periods as did the flux from a variety of optical emission lines. Figure \ref{fig:spec} shows five optical spectra (with non-uniform wavelength coverage) taken at different optical brightness states. It can be seen that during bright states, such as those seen during 2001 and 2012, lines from the Balmer series, He I $\lambda$5875, He II $\lambda$4686, [OIII]$\lambda$5007 and other highly ionized species are strong, with a moderately bright continuum dominating at longer wavelengths. In turn, during faint optical states, H$\alpha$ and other lines were very faint or absent in the spectra, with the red giant continuum prevailing toward longer wavelengths.

\begin{figure}
    \includegraphics[scale=0.57]{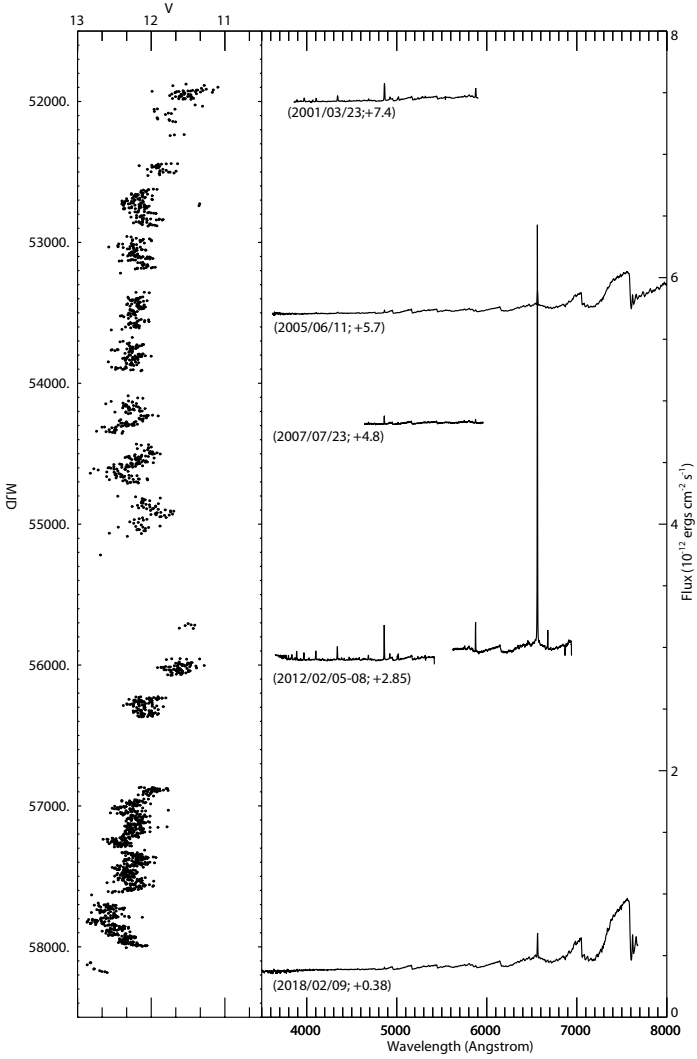}
 \caption{Comparison of optical spectra taken during faint and bright states. We indicate, in parentheses, the observation date and flux offset applied to match the observation date with the light curve in the left panel. During 2005 and 2018, well within the faint state, the red giant dominated the optical spectrum, and only a few emission lines (mostly H$\alpha$) were observed. On the other hand, in 2001 and 2012, during bright optical states, emission lines from highly ionized elements such as H, O, and Fe are strong.}
 \label{fig:spec}
 \end{figure}

\subsection{X-ray spectra.}
\label{sec:xray}

\subsubsection{\nustar~ spectral modeling and detection of reflection.}

Based on our previous experience with modeling \chandra\ data of RT~Cru and other symbiotics \citepads{2007ApJ...671..741L,2013A&A...559A...6L,2016MNRAS.461L...1M},
we fit the \nustar\ data with a spectral model that consists of a variable abundance, cooling flow modified by a complex absorber (partial covering absorption) and reflection (\texttt{constant$\times$TBabs$\times$(partcov$\times$TBabs)$\times$(reflect$\times$vmcflow\\+gauss)} in XSPEC notation).  The complex absorber in RT~Cru introduces a degeneracy with the reflection amplitude \citepads{2015ApJ...807L..30M}, and thus we modeled the \nustar\ data together with a \swift/XRT observation obtained almost simultaneously, extending the spectrum toward low energies where absorption has more influence. We allowed the cross-normalization between \nustar\ FPMA, FPMB and \swift\ to vary (notice that Figure \ref{spec:nustar} shows only the \nustar\ FPMA spectrum for clarity).

The spectrum from \nustar\ has a high enough S/N at high energies to allow us to explore the effects of reflection from the WD surface and accretion flow, and to thereby better constrain the shock temperature. Reflection produces both the 6.4 keV Fe K$\alpha$ line, observed in the spectra of many symbiotic stars, and a 10--30 keV continuum hump, which hardens the spectrum and makes the emission appear hotter. Since reflection lowers photon energy, we extended the model calculation to energies beyond the nominal bandpass of \nustar.

The plasma and reflector abundances, as well as the equivalent width of Fe K$\alpha$ 6.4 keV, influence the reflection amplitude \citepads[see e.g.,][]{1991MNRAS.249..352G,2017arXiv171009931H}. The \suzaku\ spectra indicate that the EW of the FeK$\alpha$ line is variable (see below). Spectral modeling of \chandra\ data, which are not well-suited to constrain the continuum emission, yielded a Fe subsolar abundance of 0.3$_{-0.28}^{+0.15}\times$Fe$_{\odot}$ \citepads{2007ApJ...671..741L}. In spectral models that included reflection, we linked the plasma Fe abundance to that of the reflection component, while we set the abundance of other elements to solar values in both spectral components. Finally, we also explored the dependency of the fit parameters such as $kT$ over the viewing geometry ($\cos{\mu}$) and found that $kT$ remains between the range determined above even if $\cos{\mu}$ changes to 0.94 (21.6 degrees), while the reflection amplitude decreases to $\sim$0.25. The spectra and the model with and without reflection are displayed in Figure \ref{spec:nustar} and the resulting parameters are listed in Table \ref{tab:fits}. The model without reflection, with \chisq/d.o.f.=1.12/977, yielded a maximum temperature of 64$\pm$5 keV.  The shock temperature in the model including reflection is $kT$=53$\pm$4 keV, with a reflection amplitude of 0.77$\pm$0.21 and \chisq/d.o.f.=1.07/976. Although, statistically, both models with and without reflection seem to be acceptable, we found that the detection of reflection is reliable. We constructed $\chi^{2}$ confidence contours maps (90, 95 and 99 \%) of each free-to-vary parameter against reflection amplitude for the model including reflection, using the \texttt{steppar} command in XSPEC, which performs a fit while stepping the value of the parameter through a given range. We varied the reflection amplitude between 0 and 2, allowing a wide dynamic range, and found that it is $\gtrsim$0.4 in all cases.

\begin{figure}

\includegraphics[scale=0.45]{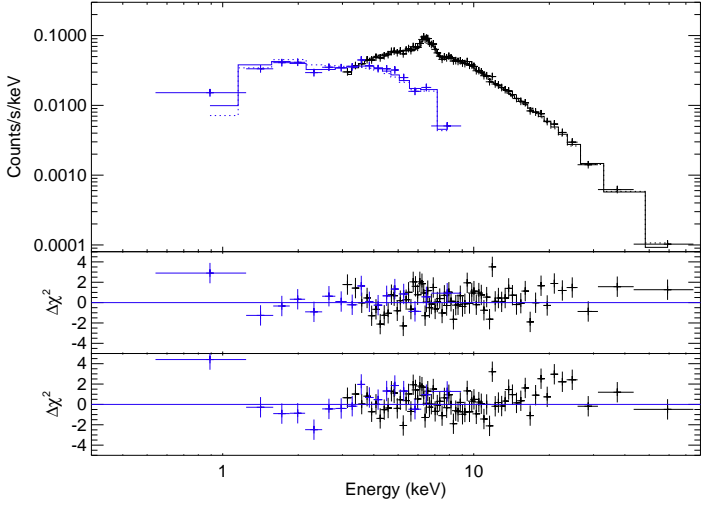}
\caption{{\em (Top):} \nustar\ ({\em black}) + \swift\ ({\em blue}) data and best spectral model (solid lines), which consists of a variable abundance, isobaric cooling flow modified by a complex absorber and reflection (see Table \ref{tab:fits} for resulting parameters). {\em Middle} and {\em bottom} panels show the fit residuals in units of \chisq\ for the model with and without reflection, respectively. }
\label{spec:nustar}
\end{figure}

\subsubsection{\suzaku\ and \swift\ spectra and variability.}

To model the \suzaku\ and \swift\ spectra, we fixed the value of $kT_{max}$ to 53 keV and used the same absorption model we used to fit the \nustar\ spectrum.  
With a single partial covering absorber, the model did not yield satisfactory results. In fact, given the strong soft X-ray variability observed (see below), it is natural to conclude that the absorption was complex, probably changing depending on the source state. Motivated phenomenologically rather than physically, we used two partial covering absorption components when modeling the \suzaku\ spectra. Modeling the \suzaku/HXD spectra with a \texttt{bremss} model to constrain any changes in the temperature structure of the boundary layer, we found that during both quiescence and bright optical states, the temperature was the same, considering the 90\% confidence error bars; $kT_{1}$=31$_{-6}^{+10}$ keV and $kT_{2}$=31$_{-5}^{+7}$ keV, respectively. The fits of the \suzaku\ XIS+HXD spectra and the residuals are shown in Figure \ref{spec:suzaku:1st}.

\begin{figure*}
\begin{center}
  \includegraphics[scale=0.3]{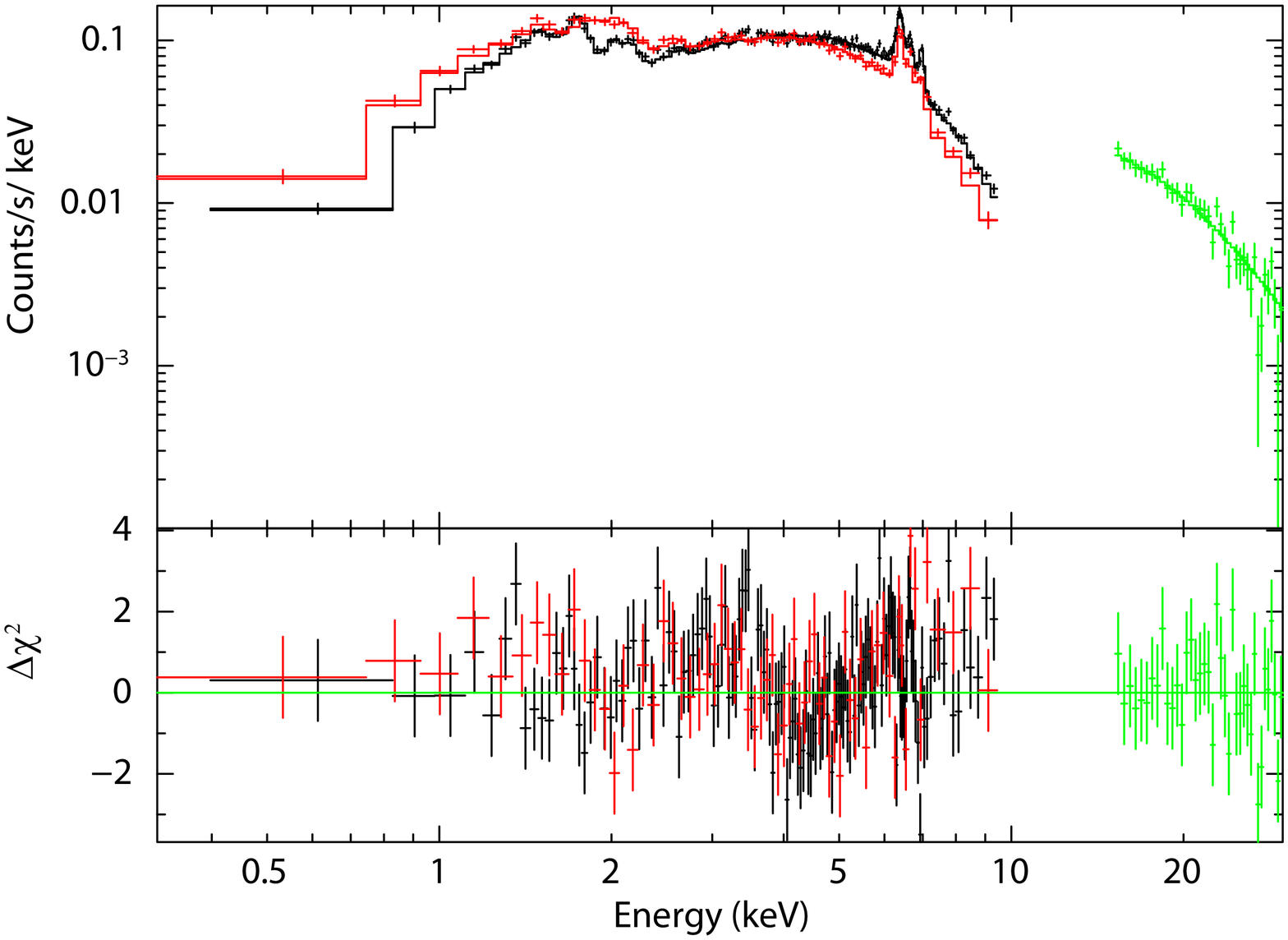}
    \includegraphics[scale=0.3]{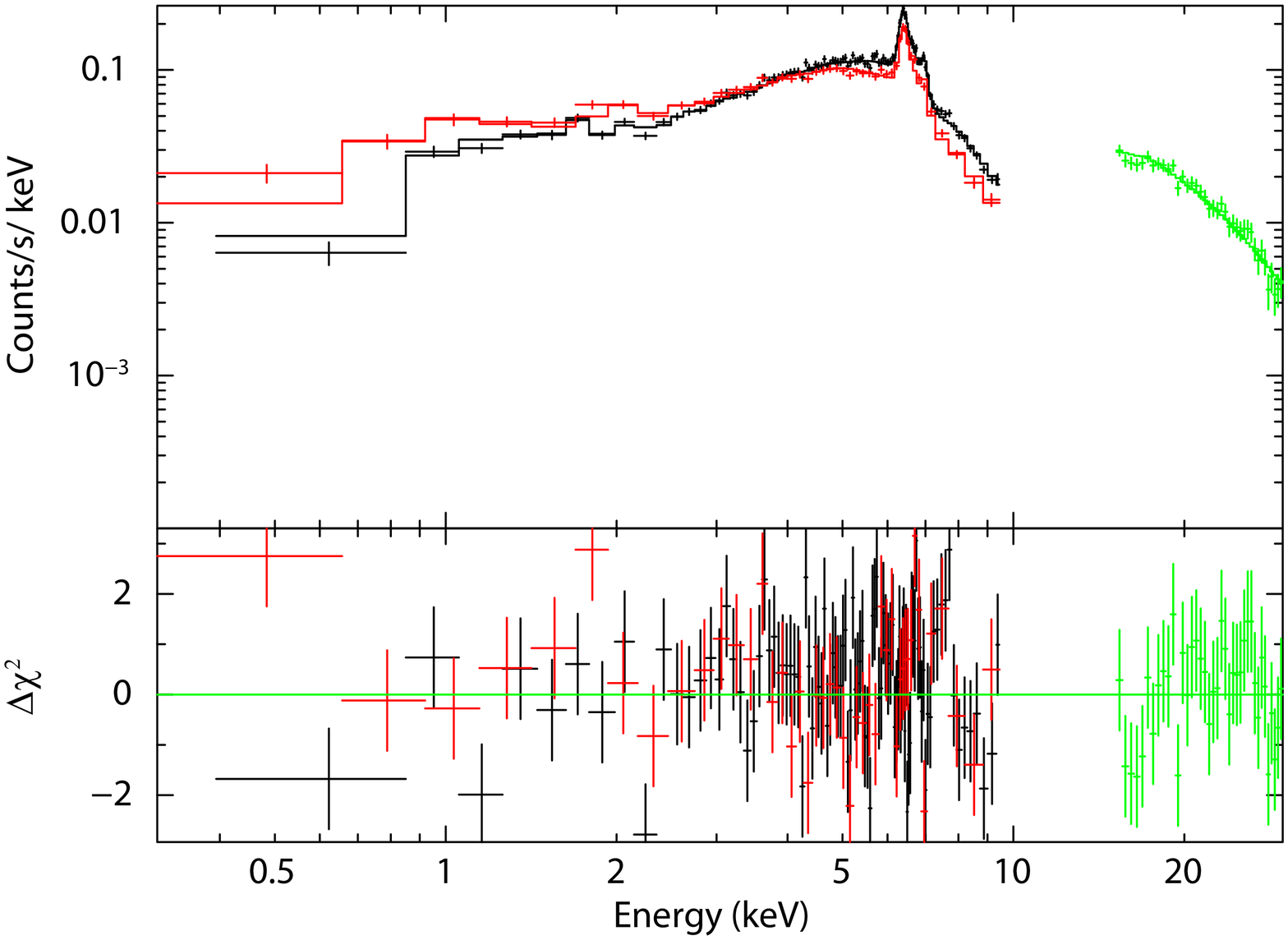}
\caption{{\it (Left):} \suzaku\ spectrum taken in 2007 (ObsID 402040010). {\it (Rigth)} \suzaku\ spectrum taken in 2012 (ObsID 906007010). XIS0+3 spectra and model are shown in black, XIS1 in red and HXD in green.  Best spectral model (solid line), consists of a variable abundance, isobaric cooling flow with fixed kT$_{max}$=53 keV, modified by two complex, partial covering absorbers (see Table \ref{tab:fits} for resulting parameters).}
\label{spec:suzaku:1st}
\end{center}
\end{figure*}

We observed strong variability in low-energy X-rays (E$\lesssim$10 keV; see Figure \ref{fig:aavso}), which, given the stability of the BAT hardness ratio, was most likely due to changes in the amount and nature of intervening absorption and/or the accretion rate onto the WD.  The hardness ratio of the soft X-ray emission (HR; 5.5-10.0~keV/0.3-5.5~keV) changed significantly only on time scales of days to years, as observed with \swift/XRT, whereas the same ratio remained approximately constant on the shorter timescales observed with \suzaku\ (Figure \ref{lc:varia}).  The \suzaku\ spectra indicate that the EW of the FeK$\alpha$ line was also variable, with values of 230 eV (ObsID 402040010) and 350 eV (ObsID 906007010).  

\begin{figure*}
\begin{center}
 \includegraphics[scale=0.8]{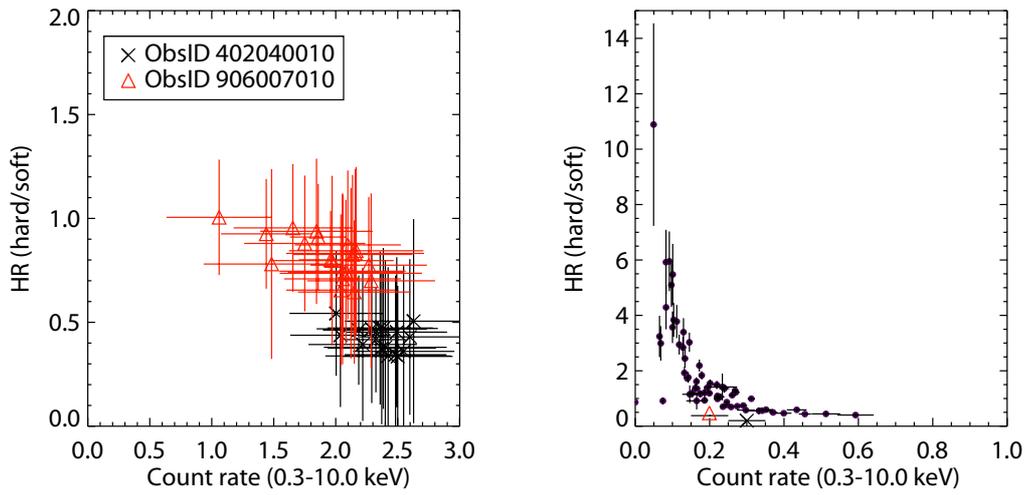}
\caption{{\em (Left):} Hardness ratios (5.5-10/0.3-5.5 keV) as a function of flux for the \suzaku\ data derived from each satellite orbit during both observations. {\em (Right):} Same hardness ratios vs. flux but for each \swift/XRT observation and both \suzaku\ observations, which were folded through \swift/XRT response in order to obtain the equivalent count rates. }
\label{lc:varia}
\end{center}
\end{figure*}

 \begin{landscape} 
 \begin{center}
 \begin{table} 
\caption{\nustar + \swift\  and \suzaku\ X-ray spectral fitting results. X-ray flux and luminosity, in units of 10$^{-13}$ ergs s$^{-1}$ cm$^{-2}$ and 10$^{31}$ ergs s$^{-1}$, respectively, are calculated in the 0.3-80.0 keV energy band. Iron abundances in units of abundances from \citetads{2000ApJ...542..914W}. Luminosities and $\dot{M}$ are determined assuming a distance of 2 kpc.}
\label{tab:fits} 
 \begin{small}     
 \begin{tabular}{|l|c| c|c| c| c| c |c |c |c |c| c |}   
          \hline
          \multicolumn{12}{|c|}{\texttt{constant$\times$TBabs$\times$(partcov$\times$TBabs)$\times$(partcov$\times$TBabs)$\times$(vmcflow+gauss)}}\\ 
 \hline       
\suzaku & Count rate & $n_{H}$ & Covering   & kT$_{max}$ &Fe/Fe$_{\odot}$ & $\lambda_{Fe K\alpha}$ & EW (Fe K$\alpha$) & F$_{X}$ & L$_{X}$ & $\dot{M}$ &$\chi^{2}_{\nu}$/dof \\
 &  &[10$^{22}$ cm$^{-2}$] &  Fraction & keV & & keV  & eV & &  & 10$^{-9}$ M$_{\odot}$ yr$^{-1}$ &\\
 \hline
402040010 & XIS0+3:0.65 & 0.12$\pm$0.07\tablefoottext{a}&      & 53        &   0.67$\pm$0.01    & 6.39$\pm$0.01& 230$\pm$10 & 1091$\pm$10 & 5220$\pm$50 & 3.7$\pm$0.1&1.04/2116 \\
           & XIS1: 0.62  & 1.4$\pm$0.1\tablefoottext{b} & 0.87$\pm$0.03 &       &     &                &  &          &    &  &  \\
           & HXD: 0.13   & 12$\pm$1\tablefoottext{b}& 0.60$\pm$0.02 &              &                &       &        &    &  &  &\\
           
906007010 & XIS0+3:0.61 & 0.06$\pm$0.04\tablefoottext{a} &      & 53      & 0.48$\pm$0.06 & 6.41$\pm$0.01 & 350$\pm$10 & 1983$\pm$14 & 9486$\pm$60 &6.7$\pm$0.3&1.05/1660 \\
          & XIS1:0.54   & 4.8$\pm$0.4 \tablefoottext{b}& 0.90$\pm$0.01 &                &                & & &  &&  &\\
          & HXD:0.21    & 28$\pm$3\tablefoottext{b}& 0.74$\pm$0.02 &                &                & & &  && & \\ 
\hline
          \multicolumn{12}{|c|}{\texttt{constant$\times$TBabs$\times$(partcov$\times$TBabs)$\times$(vmcflow+gauss)}}\\ 
\hline
 & Count rate & $n_{H}$ & Covering Fraction & kT$_{max}$ &Fe/Fe$_{\odot}$ & \multicolumn{3}{|c|}{F$_{X}$} & L$_{X}$ & $\dot{M}$ &$\chi^{2}_{\nu}$/d.o.f. \\
 \hline
\nustar\ + \swift & FPMA:0.55 &1.7$\pm$0.2 \tablefoottext{a} &    &   64$\pm$5   & 0.44$\pm$0.07 & \multicolumn{3}{|c|}{1376$\pm$10} & 6582$\pm$50 & 4.3$\pm$0.4 & 1.12/977 \\
& FPMB: 0.59  & 29$\pm$5   & 0.51$\pm$0.03       &              &        & \multicolumn{3}{|c|}{}                   &    &   &\\
& XRT: 0.19   & &       &              &         &  \multicolumn{3}{|c|}{}                    &    & &   \\           
\hline
\multicolumn{12}{|c|}{\texttt{constant$\times$TBabs$\times$(partcov$\times$TBabs)$\times$(reflect$\times$vmcflow+gauss)}}\\ 
\hline
  & Count rate & $n_{H}$ & Covering Fraction   & kT$_{max}$ &Fe/Fe$_{\odot}$ & \multicolumn{2}{|c|}{Ref. Amplitude} & F$_{X}$ & L$_{X}$ & $\dot{M}$ &$\chi^{2}_{\nu}$/d.o.f. \\
 \hline
&  &1.3$\pm$0.3 \tablefoottext{a} &   &   53$\pm$4 & 0.46$\pm$0.06 & \multicolumn{2}{|c|}{0.77$\pm$0.21}  & 1030$\pm$7 & 4925$\pm$30 & 4.1$\pm$0.2 & 1.07/976 \\
&  & 12$\pm$5   & 0.6$\pm$0.1    &              &        & \multicolumn{2}{|c|}{}         &          &    &   &\\
\hline                  
\end{tabular}
\end{small}
\end{table}
\tablefoottext{a}{Absorption column of the \texttt{TBabs} component.} 
\tablefoottext{b}{Absorption column of the \texttt{(partcov$\times$TBabs)} component.}

\end{center}
\end{landscape}

 \subsection{Oscillation amplitude upper limits and flickering}
 \label{flickering}

 Within every dataset analyzed here, we searched for periodic modulation. Specifically, we computed the Fourier power spectrum and fit a simple power-law model to the region dominated by red noise, then looked for peaks in the power spectrum that would exceed the detection threshold \citepads[a detailed description of the method can be found in][]{2005A&A...431..391V}. In the case of \nustar\ and \suzaku\ data, we searched in the frequency range 2$\times$10$^{-5}$ Hz $<$ $f$ $<$ 0.008 Hz and in the case of \swift/UVOT in the frequency range 0.001 Hz $<$ $f$ $<$ 0.1 Hz. We did not find peaks in any power spectrum exceeding our detection limits.  The \nustar\ power spectrum is displayed in Figure \ref{fig:vaughan} where we did not detect peaks with power greater than the 3$\sigma$ detection limit. Figure \ref{fig:vaughan} also includes a panel with the sinusoidal fractional amplitude \citepads[as detailed in][]{1996ApJ...468..369I} as a function of frequency for the 3$\sigma$ upper limit power spectrum. 
 
\begin{figure}
\includegraphics[scale=0.5]{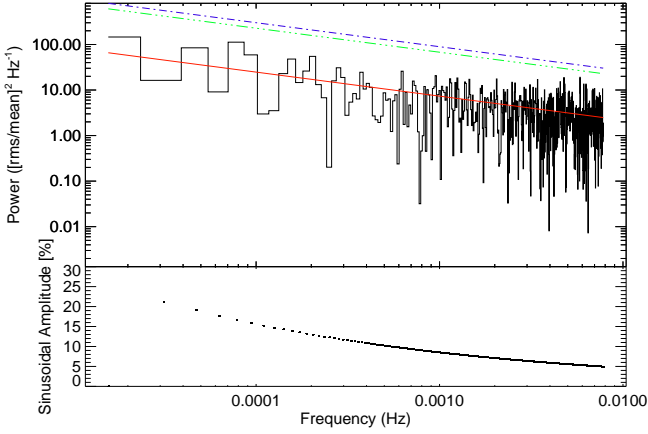}
\caption{{\em (Top)}: \nustar\ power spectrum from an FPMA+FPMB light curve with bins of  64 s. The power law model for the red noise is shown in solid (red) line and the 95 (green) and 99.74 (blue) per cent upper limits on the expected power are shown in dotted-dashed curves.  {\em (Bottom)}: Upper limit (3$\sigma$) on the sinusoidal fractional amplitude as a function of frequency.}
\label{fig:vaughan}
\end{figure}
 
{\em Swift}/UVOT data taken in event-mode allowed us to study, for the first time in RT~Cru, the UV flickering on the seconds-to-minutes time scales.  During each
snapshot taken in UVOT event mode, we detected strong variability with fractional amplitudes reaching values of 100 \% in light curves binned at 10 s
(see example in Figure \ref{lc:uvot-event}). From the light curve of each event-mode snapshot we derived the fractional $rms$ amplitude, which is plotted in Figure \ref{fig:sspex_time} versus observing time.  The X-ray light curves also showed strong aperiodic variability, with $rms$ values from 0.10 (\nustar) to 0.29(\suzaku/XIS/906007010), consistent with previous findings \citepads[see][]{2007ApJ...671..741L}. \nustar\ and \suzaku\ light curves, with a bin size of 600 s, are displayed in Figure \ref{fig:xlc}.   

\begin{figure}
\includegraphics[scale=0.5]{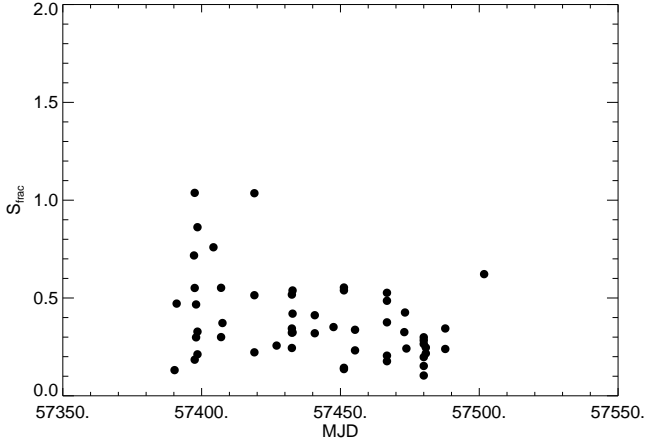}
\caption{\swift/UVOT event-mode variability on time scales of seconds. The plots show the fractional $rms$ amplitude versus observing date for those observations taken in UVOT event mode.}
\label{fig:sspex_time}
\end{figure}

\begin{figure}
\includegraphics[scale=0.5]{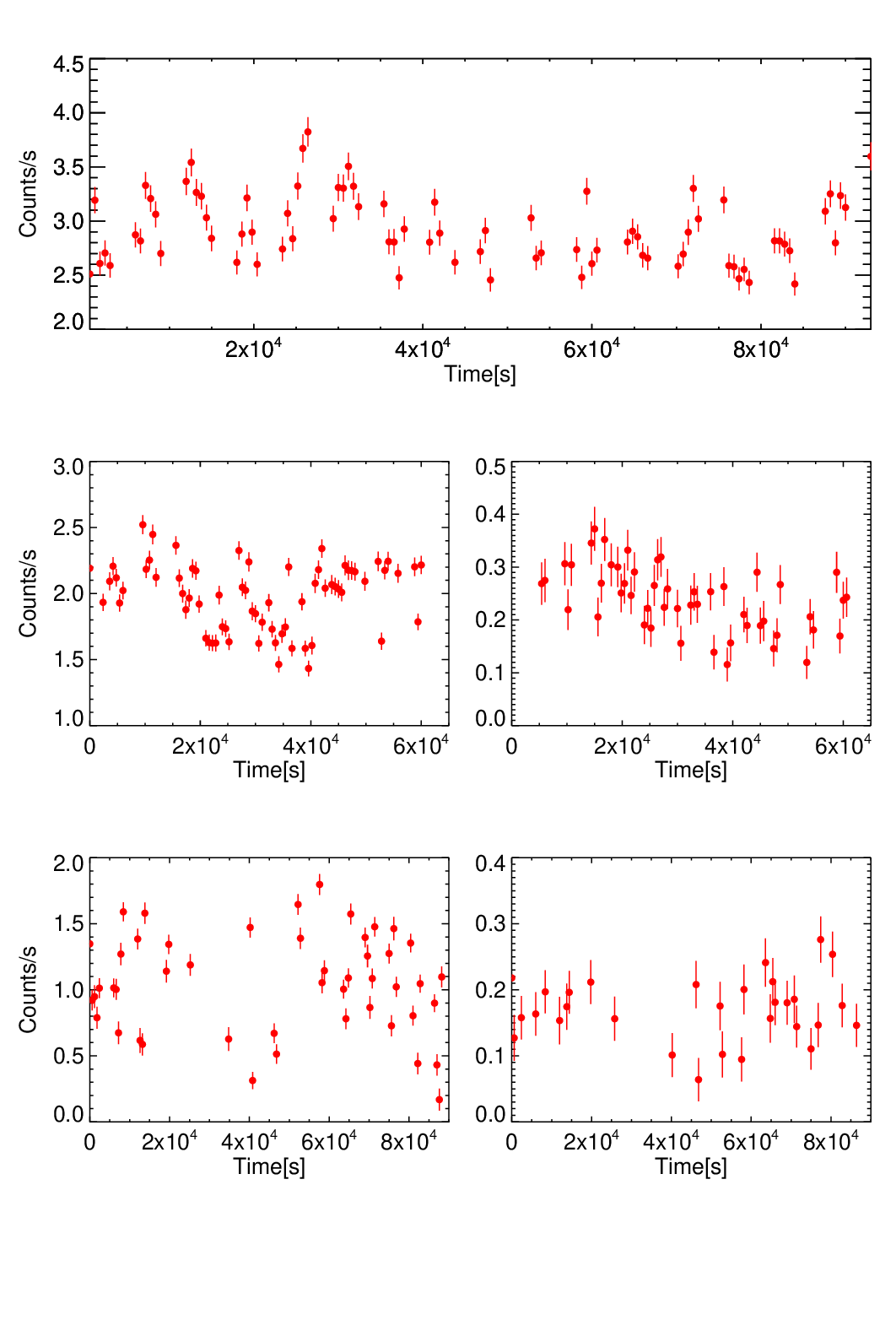}
\caption{X-ray light curves with 600 s bin size. $(a)$ \nustar\ FPMA+FPMB ($s_{frac}$= 10\%); {\em (b)} \suzaku/XIS ObsID 402040010 ($s_{frac}$=13\%); {\em (c)} \suzaku/HXD ObsID 402040010 ($s_{frac}$= 26\%); {\em (d)} \suzaku/XIS ObsID 906007010 ($s_{frac}$= 36\%); \suzaku/HXD ObsID 906007010 ($s_{frac}$= 29\%).}
\label{fig:xlc}
\end{figure}

\section{Discussion.}
\label{sec:disc}

\subsection{Evidence for boundary-layer origin of X-ray emission}

The first high-S/N X-ray data of RT~Cru taken with \chandra/HETG indicated that the spectrum and light curve were compatible with those emanating from an accretion disk boundary layer \citepads{2007ApJ...671..741L} around a massive white dwarf (M$_{WD} \gtrsim$1.3 M$_{\odot}$).  \citetads{2007ApJ...671..741L} estimated the mass of the WD assuming that the boundary layer was optically thin and therefore the plasma maximum temperature was set by the mass of the WD. Recently, \citetads{2016A&A...592A..58D} claimed that the value of $\dot{M}$ found from the modeling of \integral\ + \swift/XRT and \suzaku\ data would require a mostly optically {\em thick} boundary layer in the theoretical framework derived by \citetads{1995ApJ...442..337P} and \citetads{2014A&A...571A..55S}; they therefore proposed an alternative scenario in which the X-ray emission is produced in a magnetically channeled flow, analogous to cataclysmic variables of the intermediate polar type and constrained the WD mass to M$_{WD}$= 0.9--1.1 M$_{\odot}$. 

Our observations, however, are difficult to explain in the magnetic-channeled accretion scenario, where X-rays arise in the post-shock region of an accretion column. Our data present multiple pieces of evidence against this picture: 

\begin{itemize}

\item[$\bullet$] \nustar\ data constrain the fractional amplitude of any X-ray oscillation due to magnetic accretion to $\lesssim$5 \% for periods of about $\sim$2 mins
while the sensitivity rapidly decreased toward shorter frequencies due to the presence of red noise (see Figure \ref{fig:vaughan}). Previously, a $\sim$50 ks \chandra~observation constrained the amplitude of periodic oscillations to be less than 8\% for periods between 30 s and 12 mins \citepads{2007ApJ...671..741L}, and our analysis supports the lack of a periodic modulation with additional upper limits.  

\item[$\bullet$] In the case of magnetically channeled accretion, the temperature at the shock depends on the height at which the shock is produced and the internal radius of the accretion disk, that is, $kT_{max}=\frac{3G\mu m_{p} M_{WD}}{8}\left(\frac{1}{R_{WD}+h_{sh}}-\frac{1}{R_{in}}\right)$ \citepads{2005A&A...435..191S}. The inner radius of the accretion disk, R$_{in}$, in magnetic CVs is the region where the magnetic field pressure and the accretion ram pressure are in balance. An increase in the accretion rate would break that balance and reduce the distance between the white dwarf surface and the inner region of the accretion disk, as possibly happened in the most recent dwarf-novae-type outburst of the intermediate polar GK~Per \citepads[see ex.]{2004MNRAS.352.1037H,2005A&A...439..287V,2017MNRAS.469..476Z} where the shock temperature decreased from 26 to 16 keV during the last dwarf-novae-type outburst. If we saw kT$_{max}$ change in such a way as to follow Suleimanov's equation, that would have been evidence for a magnetic nature; we did not see such a change.

As discussed above, the non-variable \swift/BAT hardness ratio indicates that the shock temperature did not change significantly, with the \suzaku/HXD spectra characterized by temperatures during quiescence and brightening optical states of $kT_{1}$=31$_{-6}^{+10}$ keV and $kT_{2}$=31$_{-5}^{+7}$ keV, respectively. Therefore, if accretion onto the WD in RT~Cru is magnetically driven, either the magnetic field is strong enough to ensure that the magnetosphere is located far enough from the WD surface that a significant increase in the accretion rate does not imply a detectable change in the shock temperature or the magnetic field intensity is low enough that the inner radius of the accretion disk is close to the white dwarf and the accretion rate per unit area does not increase significantly with an increase in the accretion rate.  

\item[$\bullet$] As in magnetic CVs, the accretion disk is truncated at the magnetosphere and most of the gravitational energy of the accretion flow is radiated in the accretion column after the shock. It is therefore expected that the ratio of UV-to-X-rays flux is lower than in non-magnetic systems, where the disk extends all the way down to the WD surface. For example, we analyzed \swift\ XRT+UVOT observations of the intermediate polars EX~Hya (ObsID 00037154002) and GK~Per (ObsID 00030842075) in quiescence and outburst, respectively, and in both cases the $\frac{F_{UV}}{F_{X}}$ ratio was lower than 1 (0.63 and 0.60 for EX~Hya and GK~Per, respectively). In turn, in RT~Cru, this ratio is around 1 most of the time (see Figure \ref{fig:uv_to_xr}).

Our modeling of \nustar\ and other datasets therefore strongly suggest that the X-ray emission in RT~Cru originates in the accretion disk boundary layer around a non-magnetic WD.  The strong flickering detected (see Section \ref{flickering}) in X-rays and UV is also a feature that RT~Cru shares with symbiotics, such as RS~Oph, T~CrB, MWC~560, Z~And, V2116~Oph, CH~Cyg, o~Cet, EF~Aql, V648 Car and likely V407~Cyg \citepads{2017AN....338..680Z}; all of them powered by accretion through a disk.

\end{itemize}

\subsection{The optical depth of the boundary layer.}

Depending on the accretion rate, the boundary layer can be optically thin or thick to its own radiation.  Here we describe how our observational diagnostics reveal the boundary layer in RT~Cru to be typically optically thin. 

In a slow rotating white dwarf, half of the accretion luminosity is expected to be released in the X-ray regime from the boundary layer and the other half to be radiated from the Keplerian accretion disk in UV and optical. In quiescence, when the accretion rate is low enough \citepads[less than  about 10$^{-9.5}$ to 10$^{-9}$ \ms yr$^{-1}$][]{1995ApJ...442..337P} the boundary layer is expected to be optically thin to its own radiation and therefore the ratio UV+optical/X-ray luminosities should be equal to 1. In contrast, during high- accretion-rate episodes, such as those in dwarf novae in outburst, or persistent high-accretion regimes as in novae-like, the boundary layer often becomes optically thick to its own radiation, and the peak of its emission shifts toward longer wavelengths, causing the UV+optical/X-ray luminosity ratio to be significantly greater than 1.  

Empirically, a few authors \citepads[see ex.][]{2005ApJ...626..396P} have used simultaneous X-rays and UV observations and test the ratio of UV-to-X-ray luminosities to trace the optical depth of the boundary layer. During the main part of the outburst of the dwarf nova SS~Cyg, L$_{opt}$/L$_{RXTE}$ = 100, and L$_{EUVE}$/L$_{RXTE}$ = L$_{disk}$/L$_{BL}$ $>$ 3000 \citepads{2003MNRAS.345...49W}. \citetads{2014ApJ...794...84B} studied \swift\ observations of three novae-like cataclysmic variables, V592~Cas, MV~Lyr, and BZ~Cam, for which we found that L$_{UVW1}$/L$_{X} \gtrsim$ 10, even before UV extinction correction. \citetads{2005ApJ...626..396P} studied a sample of ten dwarf novae, with nine of them being in quiescence and derived UV luminosities from single, wide-band filter observations with the Optical Monitor on board XMM-$Newton$. They found that L$_{OM}$/L$_{X}$ was around 1 for all the systems in quiescence. The values of L$_{UVW1}$/L$_{x}$ we measured were closer to those found by \citetads{2005ApJ...626..396P} for accreting white dwarfs with optically thin boundary layers than to the values of $>$10 to 100 found in accreting white dwarfs with optically thick boundary layers.

As we have collected data from the \swift\ XRT and UVOT simultaneously (only single filter observations), we want to test if the ratio of X-ray-to-UV flux derived from these data can be used as a proxy for the optical depth of the boundary layer, but first we need to quantify the need for a bolometric correction in the UV single-filter observations. To measure the disk luminosity is complicated in CVs and is even more so in symbiotics, where both the nebulae and red giant can contribute appreciably to the UV-optical flux.  In order to quantify the need for a bolometric correction in the single-filter UVOT observations, we can estimate the bolometric flux from the Keplerian portion of the accretion disk by:  

\begin{equation}
F_{disk}=\frac{G M_{WD} \dot{M}}{8 \pi R_{WD} d^2}, 
\end{equation}

where $\dot{M}$=4.1$\times$10$^{-9}$ \ms\ yr$^{-1}$ is the mass transfer rate in the disk, equal to the accretion rate measured while fitting the \nustar\ data, $M_{WD}$=1.25 \ms\ is the white dwarf mass, $R_{WD}$=7.1$\times$10$^{8}$ cm is its radius \citepads[using the M-R relation from][]{2005A&A...441..689A} and $d$=2 kpc is the distance. The bolometric flux is therefore $\sim$6$\times$10$^{-11}$ \fluxcgs.

The average observed flux through each filter is: $\overline{U}$=2.2$\times$10$^{-11}$ \fluxcgs; $\overline{UVW1}$=2.2$\times$10$^{-11}$ \fluxcgs; $\overline{UVM2}$=4.7$\times$10$^{-11}$ \fluxcgs\ and $\overline{UVW2}$=6.4$\times$10$^{-11}$ \fluxcgs. The single-filter measured UV flux is then commensurate with the expectation from a standard Keplerian disk and we therefore assume that the UVOT fluxes can be used to evaluate the L$_{UVOT}$/L$_{XRT}$ ratio. In Figure \ref{fig:uv_to_xr} we plot the absorption-corrected ratio of UVOT-to-X-ray (0.3-80.0 keV) fluxes from all those data taken with \swift/XRT/UVOT. This figure suggest that, apart from a few dates when F$_{UV}$/F$_{X}$ is $\gtrsim$ 1, the boundary layer is mostly optically thin most of the time. 

\begin{figure}
\begin{center}
\includegraphics[scale=0.45]{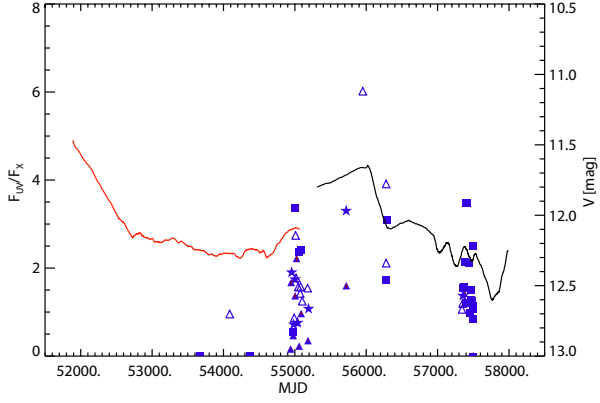}
\caption{{\em (Top):} Ratio of reddening-corrected UV flux (U: filled $\triangle$; W1:$\star$; M2: open $\triangle$; W2: $\Square$) over unabsorbed X-ray flux in the 0.3-80 keV band for each \swift\ observation. Over-plotted are the ASAS (red solid line) and AAVSO (black solid line) V-band light curve, smoothed for clarity. } 
\label{fig:uv_to_xr}
 \end{center}
\end{figure}
 
The stability of the temperature of the X-ray emitting plasma supports our contention that the boundary layer remained mostly optically thin. An X-ray spectrum alone cannot reveal the optical depth of the boundary layer because the measured temperature is that of the plasma that is already radiatively cooling.  However, unless the absorption column is very high, $N_{H}\sim$10$^{24}$ cm$^{-2}$, the hardness ratio of the hard X-ray emission observed with \swift\ BAT is a proxy for plasma temperature. None of our \suzaku\ fits indicate absorption on the order of $N_{H}\sim$10$^{24}$ cm$^{-2}$. The constancy of the hardness ratio (see Figure \ref{fig:aavso}) since the start of observations with \swift\ therefore suggests that the temperature of the boundary layer did not change significantly. The constancy of the \swift/BAT hardness ratios (Figure \ref{fig:aavso}) during the peak of optical light, as well as the L$_{UV}$/L$_{X}$ ratio, indicate that there were no significant changes in the shock temperature. We therefore assume that the boundary layer was mostly optically thin to its own radiation and that its  temperature was set by the mass of the accreting WD.  The temperature derived from \nustar\ data therefore accurately reflects the true WD mass.  

Finally, our contention that the boundary layer remains optically thin most of the time is also supported by the fact that there was not a significant decrease in the X-ray flux as optical flux increased, something that has been observed in SS~Cyg and some other DNe. Therefore, since in DNe the spectral softening seems to be associated with X-ray quenching, it is reasonable to assume that kT$_{max}$ for RT~Cru did not change significantly during optically bright events (consistent with the BAT hardness ratio being fairly constant throughout the optical brightening; see Figure \ref{fig:aavso}). 

Models of accretion disk boundary layer indicate that the transition from optically thin to thick regime should occur at $\dot{M}\gtrsim$10$^{-9}$ M$_{\odot}$ yr$^{-1}$ for a 1 M$_{\odot}$ WD  \citepads{1995ApJ...442..337P} or even lower values as proposed by \citetads{2014A&A...571A..55S}. 
The observed L$_{UV}$/L$_{X}$ ratio, together with the best fit model to \nustar\ data, indicates that the accretion rate threshold where most of the boundary layer remains optically thin can be as high as $\dot{M}\gtrsim$4.1$\times$10$^{-9}$ M$_{\odot}$ yr$^{-1}$ (albeit uncertainties in the distance). 

\subsection{First detection of X-ray reflection in a symbiotic star} 

We have detected, for the first time in a symbiotic system, a clear signature of reflection in the hard X-ray spectrum obtained with \nustar, with an amplitude of 0.77$\pm$0.21. This detection allows us to accurately determine the plasma shock temperature with a value of $kT$=53$\pm$4 keV, which in the case of strong shock conditions and optically thin boundary layer can be translated into a WD mass of 1.25$\pm$0.02 \ms. 

\subsection{Long-term, multi-wavelength changes from a variable accretion rate}

Variability on timescales of minutes --- as we observed in both the UV and X-rays --- most likely originates in the innermost region of the accretion disk. It is unlikely to be related to parameters of the binary such as orbital period or red giant pulsation, which vary on much longer timescales of days to months \citepads{2013AcA....63..405G}. Since its discovery as a hard X-ray source, it has been well established that RT~Cru is strongly variable \citepads{2010HEAD...11.1904L}. In optical, its flickering nature has been known since \citetads{1994A&AS..106..243C}. Given the amplitude of rapid, stochastic variations at optical, UV, and X-ray wavelengths, we conclude that RT~Cru is accretion powered, and that the long-term optical variability, which is accompanied by X-ray variability, implies that the accretion rate onto the white dwarf is variable. This can happen if the mass transfer rate from the donor to the accretion disk is variable, or if an instability modulates the rate at which the accretion disk transports matter inward. 

The accretion disk in symbiotics should be orders of magnitude larger than those in CVs, given their longer orbital periods. Such disks are likely to be unstable \citepads{1986A&A...163...56D,2008ASPC..401...73W}, mostly at the outer regions, which are too cool to be in the ``hot state'' and transport material efficiently. It is therefore possible that matter accumulates in the outer regions until a thermal-viscous instability is triggered and the WD accretes from the disk in a manner similar to dwarf-nova.

The optical brightening amplitude of a few magnitudes is comparable to the amplitude observed during dwarf novae-type outburst in CVs. In RT~Cru, it takes approximately 2000 days to reach maximum. If we take this time interval as the time it would take for the accumulated mass to be delivered to the WD, to be close to the local, hot-state viscous time:

\begin{equation}
t_{vis}[s] \sim 3\times 10^{5} \alpha^{-4/5} (\dot{M}/10^{16}~\ gr~\ s^{-1})^{-3/10} (M_{WD}/M_{\odot})^{1/4} (R/10^{10}~\rm cm)^{5/4}, 
\end{equation}

where $\alpha \sim$ 0.1 is the hot state viscosity parameter \citepads{2008ASPC..401...73W}, $\dot{M}$=4.1$\times$10$^{-9}$ \ms\ yr$^{-1}$ is the mass-accretion rate derived from the fit of \nustar\ data, and M$_{WD}$=1.25 \ms\ \citepads[eq. 5.69 in][]{2002apa..book.....F}. The instability could therefore develop at a distance R$\sim$10$^{12}$ cm, well within the expected size of accretion disk in these long-period binaries. 

The multiwavelength behavior in RT~Cru, however, is different from the most standard outside-in, disk instability, DNe outburst. These outbursts have a rapid rise to maximum, the time-lapse of which is determined by the distance between the region in the accretion disk where the instability has started and the boundary layer. A similar delay is observed between optical and hard X-ray emission \citepads{2003MNRAS.345...49W}. In RT~Cru, the optical brightening takes approximately 2000 days to reach maximum, but a similar time delay is not observed between optical and hard X-rays.  \suzaku/HXD and \swift/BAT observations indicate that the hard X-ray flux {\em increased in coincidence} with optical until reaching the maximum. Moreover, the decay rate after maximum is also commensurate in both optical and hard X-ray light curves.

The prototypical dwarf nova SS~Cyg exhibits a hardening of the X-ray spectrum during optically faint phases, which is explained by the low accretion rate and therefore small optical depth of the X-ray emitting plasma, while during outburst, the X-ray spectrum softens due to the increase of the accretion rate \citepads{2003MNRAS.345...49W}. Such softening of the X-ray emission is not observed during optical brightening in RT~Cru. The optical brightening around JD2451870 and the re-brightening approximately 4,000 days later could, nevertheless, have been due to disk instabilities such as those behind dwarf novae-type outbursts, if the accretion rate did not change enough to greatly increase the optical depth of the boundary layer. Finally, the fractional amplitude of the X-ray variability observed with \suzaku\ (Figure ~\ref{fig:xlc}) is higher during optical bright states (second \suzaku\ observation, ObsID 906007010); again, in the opposite sense to that observed during outburst in SS~Cyg \citepads{2003MNRAS.345...49W} where the fractional variability is stronger during the X-ray suppression (optical brightening) and smaller during the X-ray recovery (return to optical quiescence).

That the soft X-rays vary on timescales of days and longer, but not hours, suggests that changes in the amount of absorbing material have more impact on the observed 0.3-10 keV count rates than changes in the accretion rate; i.e., overall normalization.  This long-term behavior suggest that, if the absorbing material is tied to orbital motions in the accretion disk, then it is located at a relatively large distance from the X-ray source.  The first \suzaku\ observation was taken during a low-optical-brightness state, while the second observation was taken close to the optical peak (see Figure \ref{fig:aavso}). Previously we showed that the amount of absorption has changed between these two observations but also the amount of accreting material has changed, as suggested by the increase in flux in the hard X-ray energies observed with \swift/BAT and \suzaku/HXD. From the two \suzaku\ observations, we see that brightening in the hard X-ray regime is accompanied by an increase in column and covering fraction of absorption and vice-versa.  

As the optical brightness increased, there was an increase in the amount of material being dumped into the accretion disk.  But at the same time, there was a higher absorption column density that would absorb the soft energies emitted by the boundary layer. Moreover, as the L$_{UV}$/L$_{X}$ ratio increased above 1 during the optical brightening, we can speculate that the increase in the accretion rate has also changed the boundary layer optical depth but not enough to quench the hard X-ray emission. 

The mechanism/s behind the optical brightening episodes must be related to an increase in the accretion rate. The AAVSO+ASAS V-magnitude long-term light curve shows two maxima separated by approximately 4,000 days. In Figure \ref{fig:pdm} we show the phase-folded light curve with a period of 3,992 days (best period found through phase-dispersion minimization) and we include an auxiliary color scale for the observation to clearly highlight the time of phase coincidence. The fact that not only the maxima but also the fading rate after the maxima seem to repeat suggest that there could be a periodicity. However, until more data are available, this is only speculative.
If this is indeed a periodic behavior, it could be related with a long orbital period. In such a case, the correlation between optical and high-energy light curves could be naturally explained if the orbit is eccentric and the accretion rate onto the WD increases when it "travels" through the densest part of the red giant's wind.
In this case, the optical peak should not correspond to the periastron, due to the time it would take for the material to propagate through the disk.
Although most orbital periods in symbiotics are less than a thousand days, there are few exceptions, such as R~Aqr with P$_{orb}$=43.6 yrs or CH~Cyg with P$_{orb}$=15.6 yrs \citepads{2009A&A...495..931G,2009ApJ...692.1360H}.  It should be noted that using only optical photometry of RT~Cru from the ASAS survey covering the period from November 2000 to August 2009, \citetads{2013AcA....63..405G} found two peaks in the power spectrum at 325$\pm$9 and 63$\pm$1 days, and attributed these to the orbital and pulsations periods, respectively. The ASAS survey data do not cover the most recent brightening event in 2012. 

 \begin{figure*}
\includegraphics[scale=0.6]{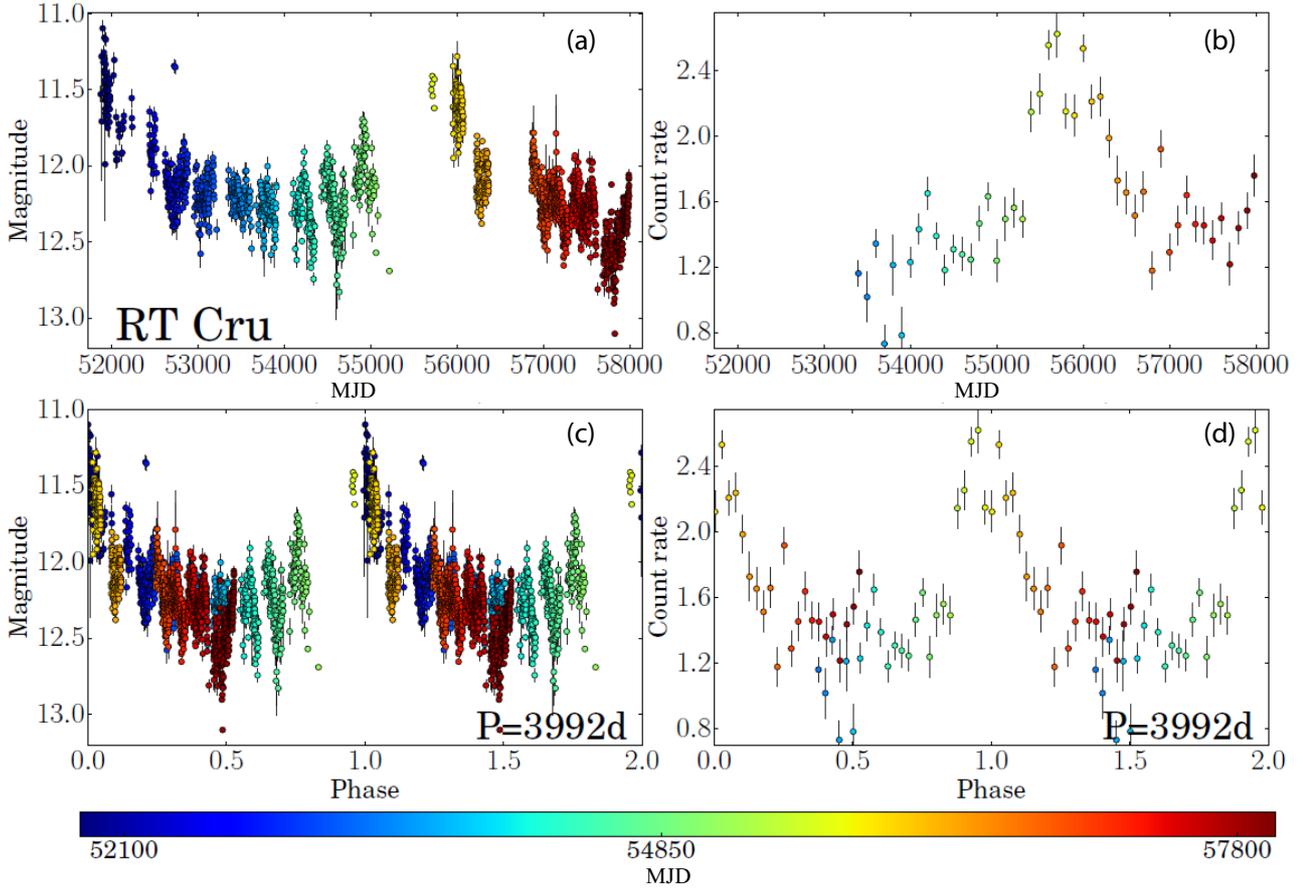}
\caption{{\em (a)}: AAVSO+ASAS V magnitude light curve. {\em (b)}: \swift/BAT light curve in 10$^{-4}$ count s$^{-1}$, with 100 days bin size. {\em (c)} AAVSO+ASAS V-magnitude light curve folded at a period of 3,992 days. {\em (d)} \swift/BAT light curve at a period of 3,992 days. We include an auxiliary color scale to clearly highlight the time of phase coincidence. The observations cover a single cycle, which is not enough to obtain a secure determination of its likelihood. However, the fact that the maxima and the fading rate seem to repeat, suggests that the periodicity could be real.}
 \label{fig:pdm}
 \end{figure*}

\section{Conclusions}
\label{sec:concl}

Throughout a multiwavelength study of the symbiotic binary RT~Cru we have gained significant insights into the accretion mode and its long-term multiwavelength behavior. We summarize our findings as follows. There is now sufficient observational evidence in support of the presence of an accretion disk that reaches the WD surface in RT~Cru, that is, the WD is not strongly magnetized, something that would disrupt the accretion disk.
 
The results from the spectral analysis indicate that a boundary layer could remain mostly optically thin if the accretion rate is higher than predicted by theoretical models.
 
Although the optical long-term variability, which shows two optical brightening events separated by $\sim$4000 days, presents few similarities to dwarf-novae disk-instability outbursts, the multiwavelength data also exhibit a few differences which cannot be explained by the dwarf-novae outburst scenario. These events are not exactly alike, but the different physical size of the disks have the potential to be able to explain the differences.
 
An alternative explanation for the long-term variability involves an enhancement of the accretion rate as the WD travels through the red giant wind in a wide orbit, with a period of about $\sim$4000 days. If the brightenings are periodic and due to orbital modulation, we expect the next brightening episode to be within the following 6 years ($\sim$ 2023), with a very similar shape during rising and fading. On the other hand, although dwarf-novae-type outbursts are likely to be quasi-periodic, the shape of the next brightening could be very different from that of the previous one. Multiwavelength observations during the next few years will be crucial to disentangle these two possible scenarios.

\begin{acknowledgements}

We thank Steven N. Shore for his comments, which helped to improve the quality of the manuscript.
We acknowledge with thanks the variable star observations from the AAVSO International Database contributed by observers worldwide and used in this research.
We thank Frank Marshall for his help with the UVOT data analysis and the entire \swift\ team for accepting and planning our multiple Target-of-Opportunity requests. We thank Simon Vaughan for his help with light curve analysis and period searches. This research has made use of the \nustar\ Data Analysis Software (NuSTARDAS) jointly developed by the ASI Science Data Center (ASDC, Italy) and the California Institute of Technology (Caltech, USA). This research has made use of the XRT Data Analysis Software (XRTDAS) developed under the responsibility of the ASI Science Data Center(ASDC), Italy. This work made use of data supplied by the UK \swift\ Science Data Centre at the University of Leicester. GJML and NEN are members of the CIC-CONICET (Argentina) and acknowledge support from grant ANPCYT-PICT 0478/14. GJML also acknowledges support from grants PIP-CONICET/2011 \#D4598 and NASA NNX10AF53G. MJA  acknowledges the financial support from CONICET in the form of Post-Doctoral Fellowship. JLS acknowledges support from grant NNX15AF19G. FMW acknowledges support for the SMARTS Observatory from the office of the Provost of Stony Brook University. The SMARTS Observatory is operated by the SMARTS Consortium.
\end{acknowledgements}

\bibliographystyle{aa}   
\bibliography{listaref_MASTER}

\end{document}